\documentclass[preprint,aps,showpacs]{revtex4}       
\usepackage{graphicx}
\newcommand{\be}{\begin{equation}}
\newcommand{\ee}{\end{equation}}
\newcommand{\bea}{\begin{eqnarray}}
\newcommand{\eea}{\end{eqnarray}}
\newcommand{\Tr}{{\rm Tr}}
\newcommand{\cL}{{\cal L}}
\begin{document}
\title{Nuclear and Neutron Star Radii}
\author{S. Schramm}
\email{schramm@theory.phy.anl.gov} \affiliation{Argonne National
Laboratory, 9700 S. Cass Avenue, Argonne IL 60439, USA}
\date{\today}

\begin{abstract}
We investigate the correlation between nuclear neutron radii
and the radius of neutron stars. We use a
well-established hadronic SU(3) model based on chiral symmetry
that naturally includes non-linear vector meson and scalar meson - vector meson
couplings. The relative
strengths of the couplings modify the nuclear isospin-dependent interactions.
We study the dependence of nuclear and neutron star
radii on the coupling strengths. The relevance of the results for parity-violating
electron-nucleus scattering and the URCA process in neutron stars is discussed.
\end{abstract}
 \maketitle

\section{Introduction}
In light of the new and planned radioactive beam facilities the
study of neutron-rich nuclei has received increased attention in recent years.
Completing the chart of metastable nuclei towards the proton and neutron drip lines
is the major goal of the endeavor with important information for
the stellar nucleosynthesis.
From a theoretical point of view nuclear calculations
toward the neutron drip-line are still quite problematic in part due to the rather poor knowledge
of the isospin-dependence of the nuclear forces. Without going to extremely neutron-rich
nuclei there is another complementary information by studying nuclear properties and the
ultimate neutron-rich systems, i.e. neutron stars, in a combined approach.
The radius of a neutron star is sensitive
to isospin forces and over a significant part of the star the corresponding
nuclear densities have values
that also occur in atomic nuclei.

A rather sensitive measure for isospin effects in
a normal nucleus is its neutron radius compared to the proton radius. In an upcoming
experiment at Jefferson Laboratory an accurate measurement of the neutron radius of
$^{208}$Pb is planned using parity-violating electron scattering \cite{jlab}.

As a basis for studying correlations between the nuclear radii with neutron star radii
a theoretical approach has to be employed that covers nuclear structure as well as
neutron star calculations in a unified way. Since the properties of neutron stars
can change significantly by including hyperon degrees of freedom, a SU(3)-flavor
approach is a convenient choice for such an investigation.
In order to study the dependence of nuclear and neutron star radii we investigate
the variation of the radii by modifying terms that influence the isospin dependence
and nuclear skin effects. Here we follow the analysis of \cite{hor2,hor1}
where in a Walecka-type model the sensitivity of nuclear and star radii on
the coupling strengths
between isovector vector mesons and scalar and vector mesons
was investigated. As it turns out such an analysis can be done in our approach in
a very natural way.

In the following we present a calculation of the radii of neutron stars and the $^{208}$Pb proton and neutron radii
in a framework that is based on a model constructed from chiral SU$_{\rm F}$(3) arguments.
The outline of the article is as follows. First in Section II we briefly introduce the general framework of
the chiral model looking at several meson interaction terms.
Then in section III we discuss the results of the calculation, studying the dependence
of neutron star radii and nuclear neutron skins on the various coupling constants.
From those results we derive
the direct correlation between neutron skin and neutron star radius.
As a test of our model we also look at the proton skin of proton-rich argon isotopes.
Finally, we investigate the consequence for neutron star cooling via the URCA process.

\section{Theoretical Framework}

In our nuclear structure as well as in the neutron star calculation we adopt a generalized
flavor-SU(3) $\sigma-\omega$ type model.
The reason for adopting this model is two-fold. For one it includes, by construction,
a nonlinear
coupling of the $\rho$ and $\omega$ meson as well as coupling between the vector mesons
and scalar mesons, so there is no need to add additional terms to an
established model to study the type of isospin dependencies mentioned before.
Secondly, in addition to successful applications in
one- and two-dimensional nuclear structure calculations \cite{beckmann,deformed},
the same model also shows a sensible behavior at high temperatures and densities\cite{hot},
which is quite important in this discussion as the inclusion of nonlinear terms in a calculations involving matter
in extreme regions of temperature or density, has the general problem that this
can produce rather uncontrolled extrapolations. A phenomenological check of the soundness
of the model predictions at extreme values is therefore reassuring.
The model used here has been successfully tested in all those regimes.

The model includes the complete set of the lowest SU(3) multiplets of baryons and mesons.
In addition extensions to higher multiplets have been discussed \cite{ziesche}.

Restricting ourselves to those degrees of freedom relevant for the discussion in this paper
the general $SU(3)$ Lagrangian \cite{paper3}
reduces to the following structure: The interaction term of
baryons, mesons, and the photon
reads
\be
\label{Lbm}
\cL_{\mathrm int} = -\sum\limits_i\bar{B_i}
\left[g_{i_\omega}\omega_0 \gamma_0 +g_{i\rho}\tau_3\rho_0^0 \gamma_0
+\frac{1}{2}e(1+\tau_3)A_0 \gamma_0 +m_i^* \right] B_i
\ee
where now the SU(3) baryon octet $B$ is reduced
to the isospinor $\left( ^p_n \right)$.
The various coupling constants of mesons and baryons result from the $SU(3)$ structure
\cite{paper3}
and the interaction of the scalar fields follows as
\bea
\cL_0^{\mathrm{chi}}  &=&  -\frac{1}{2}k_0\chi^2(\sigma^2+\zeta^2+\delta^2)+
k_1(\sigma^2+\zeta^2+\delta^2)^2
+k_2\left(\frac{\sigma^4}{2}+\frac{\delta^4}{2}+3\sigma^2\delta^2+\zeta^4\right)\nonumber\\
&& + k_3\chi\sigma^2\zeta
-k_4\chi^4
 -\frac{1}{4}\chi^4\mbox{ln}\frac{\chi^4}{\chi_0^4}
+\epsilon\chi^4\mbox{ln}\frac{(\sigma^2-\delta^2)\zeta}{\sigma_0^2\zeta_0}
\eea
Here the various scalar fields $\sigma, \delta$, and $\zeta$ correspond to the non-strange
isoscalar ($\sigma$), isovector ($\delta$)
quark-antiquark states, and the strange anti-strange state ($\zeta$).
The field $\chi$ is the scalar, isoscalar
glueball field $\chi$ introduced in \cite{schechter,pano1}.
$\chi_0$, $\sigma_0$ and $\zeta_0$ ($\delta_0 = 0$) denote the vacuum expectation values of the fields,
which are generated via spontaneous symmetry breaking.

The term
\be
\cL_{\mathrm ESB} =
-\left(\frac{\chi}{\chi_0}\right)^2\left[x\sigma+y\zeta\right]
\ee
introduces explicit chiral SU(3) symmetry breaking. The coupling strengths are
$x=m_\pi^2 f_\pi$ and $y=\sqrt{2}m_{\mathrm K}^2 f_{\mathrm K}
-\frac{1}{\sqrt{2}}m_\pi^2 f_\pi$. $f_\pi$ and $f_K$ are the pion and kaon decay constant,
respectively.
The vector meson self-interaction terms have the structure
\be
\label{Lvec1}
\cL_{vec,NL} = -a \left(\Tr\left[V^\mu V_\mu\right]\right)^2 - b \Tr\left[\left(V^\mu V_\mu\right)^2\right]
\ee
where $V$ denotes the 3x3 matrix of the vector meson multiplet. In the case of nuclei
this reduces to the form
\be
\label{Lvec}
\cL_{\mathrm vec} = - g^4_4 \left(\omega^4+6\beta \omega^2 \rho^2+ \rho^4\right) + \ldots
\ee
with the zeroth components of the isoscalar $\omega$ and neutral isovector $\rho$ field.
The couplings $g^4_4$ and $\beta$ are related to $a,b$ of equation \ref{Lvec1} via
\be
  g^4_4 = a/2 + b ~~~,~~~ \beta = 1 -{2 \over 3 (1 + a/2b)}
\ee
In addition the vector mesons interact with the glueball field $\chi$
and the nonet of scalar mesons $\Sigma$ via
\be
\cL_{\mathrm vs} = c \chi^2 \Tr[V^\mu V_\mu] + d \Tr[\Sigma^2 V^\nu V_\mu]
\ee
which reduces to
\be
\cL_{\mathrm vs} = \left[ c \chi^2 + 2 d (\sigma^2 + \delta^2 ) \right] ( \omega^2
 + \rho^2) + \ldots
\ee
In the numerical calculation we investigate the relative strength of the two
couplings of the vector mesons to the glueball field as well as to the scalars by varying
the quantity $r_\sigma$ defined as
\be
\label{rs}
\cL_{\mathrm vs} = {\alpha \over (1-r_\sigma) \chi_0^2
+ r_\sigma \sigma_0^2} \left[ (1 - r_\sigma) \chi^2 + r_\sigma (\sigma^2 +\delta^2)
\right] (\omega^2
 + \rho^2)
\ee
where the factor $\alpha$ was determined in the general parameter fit $\chi_M$
with $r_\sigma=0$ \cite{deformed}.
$r_\sigma$ shifts the relative importance of the gluonic and quark-type scalar
fields for generating the masses of the vector mesons.
The mass of the baryons are dynamically generated via their interaction with the
scalar fields following Eq. (\ref{Lbm}), which automatically leads to density and
temperature-dependent masses. Explicitly, the baryon masses $m^*_i$ are given by
\be
m^*_i =g_{i\sigma}\sigma+g_{i\delta}\delta+g_{i\zeta}\zeta\quad .
\ee

In general there is no unique way of introducing nonlinear vector couplings
in a hadronic model. Here we introduce the simplest structures within our
framework
generating the interaction terms, in fact those terms had been introduced
well before this specific investigation of isospin effects started. The
non-linear couplings in the $\omega$ channel turn out to be
very important to obtain a reasonably soft equation of
state of symmetric nuclear matter, leading to acceptable value for the compressibility
of $\kappa < 250~$MeV. In addition, away from the
saturation density the equation of state of
neutron matter is soft enough to generate neutron stars with masses $M < 2 M_{solar}$
\cite{hanauske}.

\section{Results}
\subsection{Neutron skins}
Here we show the numerical results obtained by solving the equations of motions derived from the
Lagrangian terms discussed above in a mean field approximation. In particular we will vary the two coupling strengths
$\beta$, Eq. (\ref{Lvec}), and $r_\sigma$, Eq. (\ref{rs}) and study their influence
on various observables.

First we determine the rms radii of neutrons and protons in $^{208}$Pb.
In addition as comparison we also look at the case of
the singly magic $^{138}$Ba which might be of interest in future parity-violation
experiments.
Barium isotopes have been suggested as candidates for
atomic parity violation experiments \cite{mike} in analogy to the cesium experiments
\cite{cesium}.
In those experiments uncertain neutron distributions are a major part
of the theoretical error. The calculation of $^{138}$Ba is
supposed to give an idea of the range of uncertainty for neutron skins in barium.

\begin{figure}
\includegraphics[width = 12cm]{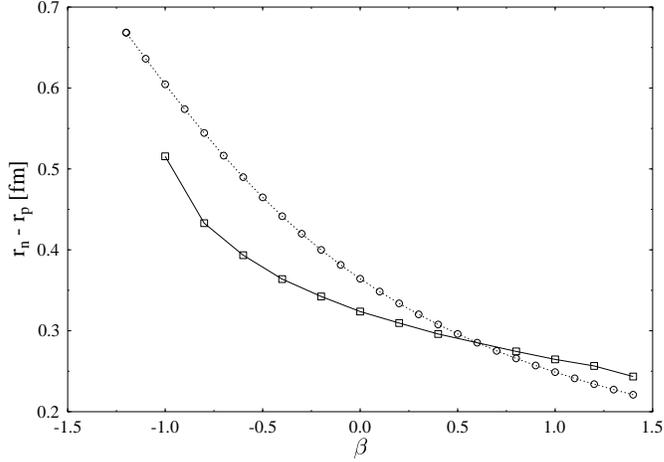}
\caption{\label{rpnbeta}Difference of neutron and proton radius in $^{208}$Pb
as function of nonlinear vector coupling strength $\beta$ with (squares)
and without (circles) refitting of parameters ($r_\sigma = 0$).
}
\end{figure}

\begin{figure}
\includegraphics[width = 12cm]{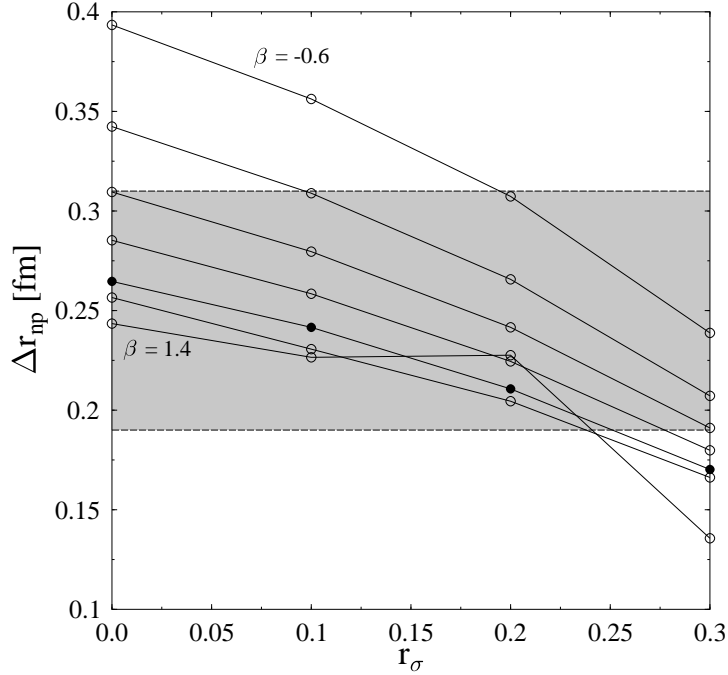}
\caption{\label{rpnall}Difference of neutron and proton radius $\Delta r_{np}$ in $^{208}$Pb
as function of relative vector-scalar meson coupling strength $r_\sigma$
for different values of $\beta = \{-0.6,-0.2,0.2,0.6,1.0,1.2,1.4\}$.
The shaded area denotes the
expected experimental accuracy for the $r_n$ measurement (tentatively) centered
at $\Delta r_{np} = 0.25\,$fm.
}
\end{figure}

The equation of motions are solved assuming spherical symmetry using the established parameter
set $\chi_M$ \cite{deformed}.
The calculation is repeated by changing the two coupling parameter $\beta$ and $r_\sigma$.
The resulting values for the difference of proton and
neutron radii $\Delta r_{np} \equiv r_n - r_p$
are shown in Figure \ref{rpnbeta}. In the variation of the parameters
we follow two general procedures.
First we use the parameters $\chi_M$
and only modify the coupling strength $\beta$ (which has a value $\beta = 1$ in $\chi_M$),
without simultaneously modifying any other parameters.
In a second run we adjust the other parameters that mainly influence the isospin behavior,
$g^4_4$ and the nucleon-rho coupling $g_{N\rho}$ to ensure a correct description of the lead charge radius
and binding energy. The result of both approaches is shown in the figure.
The general behavior is the same for both approaches. The best-fit estimate for neutron-proton
radius difference is $\Delta r_{np} = 0.25\,$fm. This number is in general agreement with other
models based on relativistic descriptions of nuclear matter \cite{ring} and is bigger than typical
Skyrme-based non-relativistic projections of $\Delta r_{np} < 0.2\,$fm \cite{skyrme}.
There is a significant dependence of $\Delta r_{np}$ on the coupling strength while varying $\beta$
between about $-1 < \beta < +1.4$, which is about the maximum range
in which one can obtain stable nuclear solutions. We will look at the parameter
range in more detail.
In order to obtain a more believable quantitative result
a refitting of the parameters is necessary, reducing $\Delta r_{np}$ by up to 20 percent.
In the following, if not noted otherwise, we will always perform a readjustment of the parameters as explained
above.
Figure \ref{rpnall} shows results for $\Delta r_{np}$ varying the coupling $r_\sigma$
for different values of $\beta$. The shaded area reflects the expected accuracy of the
neutron radius measurement of about $\pm 1\%$. Note the wiggle for the line with $\beta = 1.4$,
signalling the onset of proton and neutron rearrangements
in the solutions of the nuclear
equations that violate experimentally known nuclear charge distributions.

\begin{figure}
\includegraphics[width = 12cm]{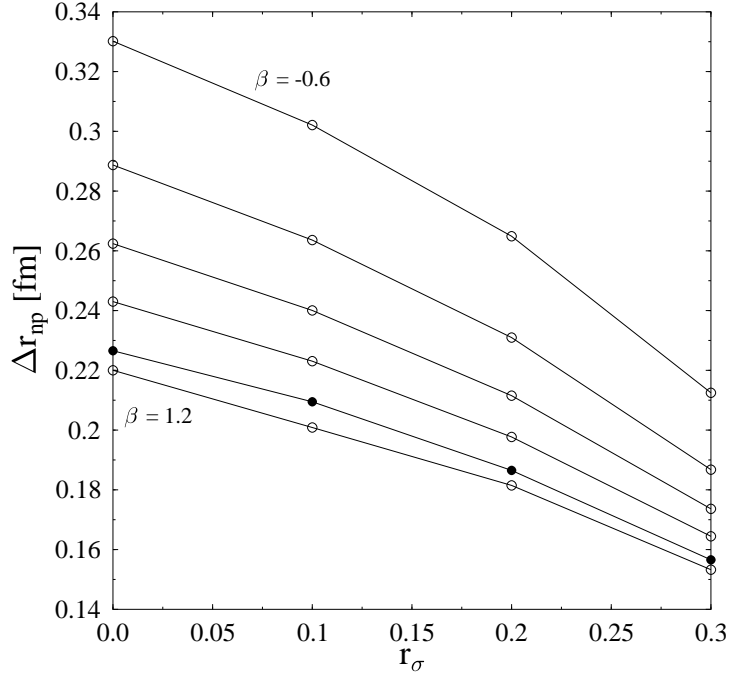}
\caption{\label{rpnBa}Difference of neutron and proton radius in $^{138}$Ba
as function of coupling strength $r_\sigma$. Curves for
different values of $\beta = \{-0.6,-0.2,0.2,0.6,1.0,1.2\}$ are shown (cf. Fig. \ref{rpnall}).
}
\end{figure}

We repeat the same calculation for the case of $^{138}$Ba, Fig. \ref{rpnBa}. The results are essentially similar.
The dependence of the neutron skin is somewhat less pronounced compared to the lead case.

\subsection{Nuclear matter}
\begin{figure}
\includegraphics[width = 12cm]{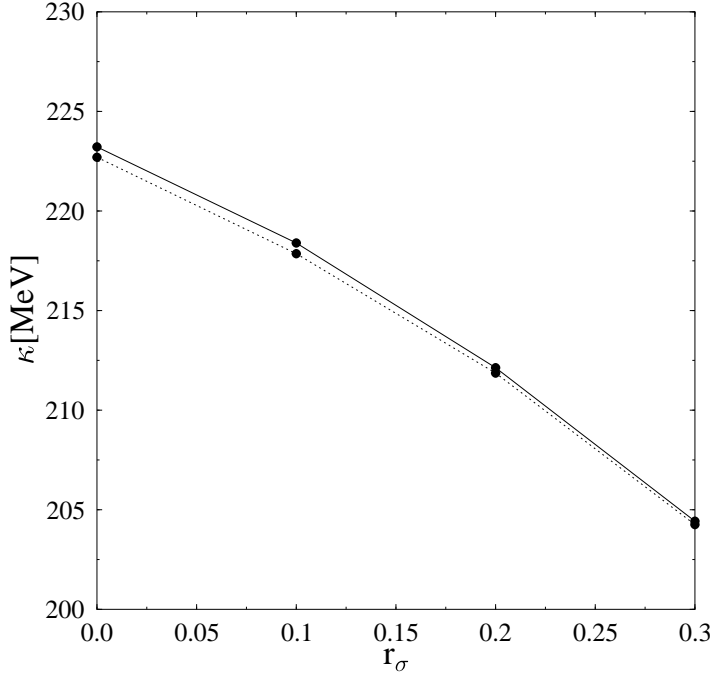}
\caption{\label{kappa}
Results for the compressibility $\kappa$ as function of the
coupling strength $r_{\sigma}$.
The two curves
show the results for the different values of $\beta = 1.2$ (full line)
and $\beta = -0.6$ (dotted line).
}
\end{figure}

In order to check that aside from a reasonable description of $^{208}$Pb
the range of parameter values yields adequate behavior of
nuclear matter at
saturation we study the values for the compressibility $\kappa$ and
the asymmetry energy $a_4$ at saturation density $\rho_0$, which has
a value $\rho_0 \sim .15\,$GeV/fm$^3$ for all studied parameter sets.
$\kappa$ and $a_4$ are defined as
\be
\kappa = 9 \rho_0^2 {\partial^2(E/A) \over \rho^2} |_{\rho=\rho_0} , \quad
\quad a_4 = \frac{\rho_0^2}{2} {\partial^2 (E/A) \over \partial (\rho_n-\rho_p)^2}
|_{\rho=\rho_0}
\ee
Fig. \ref{kappa} shows the $r_\sigma$ dependence of $\kappa$. The two curves
correspond to two extreme values of $\beta$, the other results lie between
these curves. Given these numbers one cannot exclude any parameters within
the range under consideration, all give reasonable values for the
compressibility of nuclear matter between about 205 and 225 MeV.
The results for the asymmetry energy are more restrictive.
Assuming that acceptable value for the asymmetry should be somewhere between
28 and 38 MeV, values of $r_\sigma$ larger than 0.2 yield asymmetries
$a_4$ that are too small, whereas given small couplings of scalar and vector mesons
the coupling term $\beta$ should be larger than 0. However, a quite large range
of reasonable parameter values persists.
\begin{figure}
\includegraphics[width = 10cm]{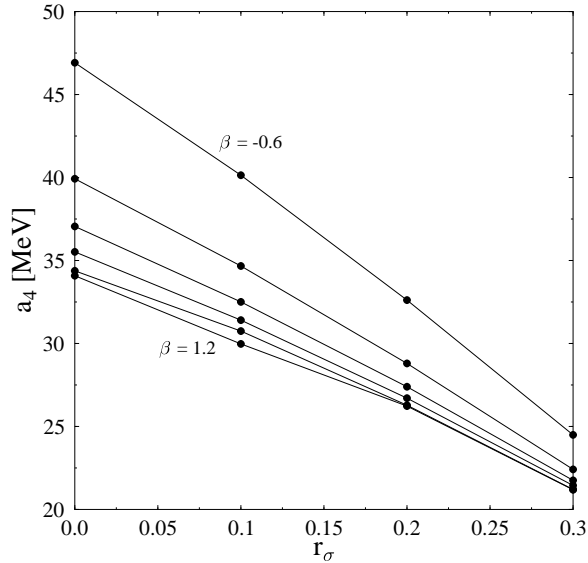}
\caption{\label{a4}
Results for the asymmetry energy $a_4$ of saturated nuclear matter
as function of the coupling strength $r_{\sigma}$.
The horizontal axis shows the value for $r_{\sigma}$ whereas the different curves are
distinguished by their values of $\beta$ ranging from $-0.6$ to $1.2$ (see
Fig. \ref{rpnBa}).
}
\end{figure}

\subsection{Neutron stars}

As a complementary source of information on the isospin-dependence of nuclear forces
a neutron star probes the extreme neutron-rich system with a content of about 90\%
neutrons of all baryons in the star.
We have performed calculations of neutron star properties in this model before
in \cite{nstar,hanauske}.
The calculation is done by integrating the Tolman-Oppenheimer-Volkov (TOV) equations
\cite{Tol} for the star
given an equation of state $\epsilon(P)$ generated from the model Lagrangian:
\begin{eqnarray}
\label{tov1} \frac{dP(r)}{dr} &= - {(\epsilon + P) (4\pi r^3 P + m) \over r^2 (1-2 m(r)/r)}
\\
\label{tov2}
\frac{d m(r)}{dr} &= 4\pi r^2 \epsilon(r)~~,
\end{eqnarray}
where $P$ and $\epsilon$ are the pressure and energy density of
the nuclear matter. $m(r)$ is the gravitational mass of the star
inside of the radius $r$. By fixing
the initial value for the pressure $P_c$ in the center of the star, with the
knowledge of the equation of state $\epsilon_c(P_c)$  one
can integrate equations (\ref{tov1}) and (\ref{tov2}) from the center of the
star outward to the
point of vanishing pressure, which defines the star radius $R$.
The calculation of $\epsilon(P)$ is discussed in detail in \cite{hot}.
The outer crust of the neutron star is modelled as
discussed in \cite{hanauske}.

\begin{figure}
\includegraphics[width = 12cm]{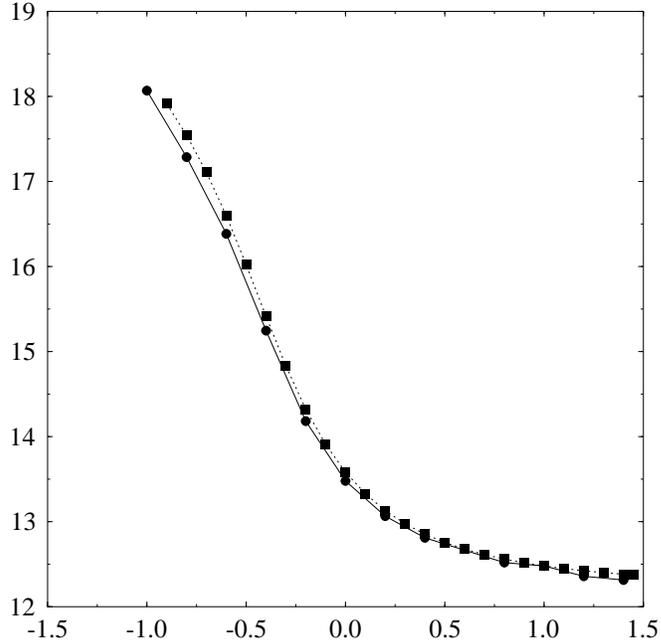}
\caption{\label{rdelta}Radius of a neutron star with $M = 1.4~M_{solar}$ varying the nonlinear
coupling strength $\beta$ for a fixed $r_\sigma = 0$.  The solid (dashed) line shows the results with (without)
refitted parameters (see text).
}
\end{figure}

Figure \ref{rdelta} shows the resulting neutron star radii for a typical neutron star with
a mass of 1.4 solar masses.
As in the nuclear case we checked for the consequences of refitting the parameters.
But in this case the refitting does not affect the neutron star properties appreciably as
can be inferred from the figure.
There is a strong dependence of the neutron star radius on
the coupling strength with a variation of about 5.5 km over the whole range of values
for $\beta$ that yield reasonable nuclei. However, with reference to the
results for the asymmetry shown in Fig. \ref{a4} in this case for $r_\sigma=0$
$\beta$ should be chosen to be larger 0, which limits the variation of the star
radius to about 1 km.
Another uncertainty entering the theoretical determination of neutron star radii
comes from the treatment of the hyperons in the star. Although our model is based
on a SU(3) model and therefore automatically contains hyperons ($\Lambda, \Sigma$
and $\Xi$ baryons, see \cite{hanauske} for a detailed discussion),
due to the poor knowledge of the behavior of hyperons in dense matter, results might
vary with the treatment of the hyperon sector. To check the limit of the uncertainties
we switch off (by hand) the hyperon densities and repeat the radius calculations for the
example of the best-fit parameter set $\chi_M$. The variation for a 1.4 solar mass star,
$R_{hyper} = 11.42~$km and $R_{nohyper} = 12.46~$km is of the same order as
the variation due to the meson-meson couplings. So in addition to a good knowledge
of the nuclear isospin forces some information on hyperons in matter is needed
to draw more reliable conclusions from neutron star radius data.

\begin{figure}
\includegraphics[width = 12cm]{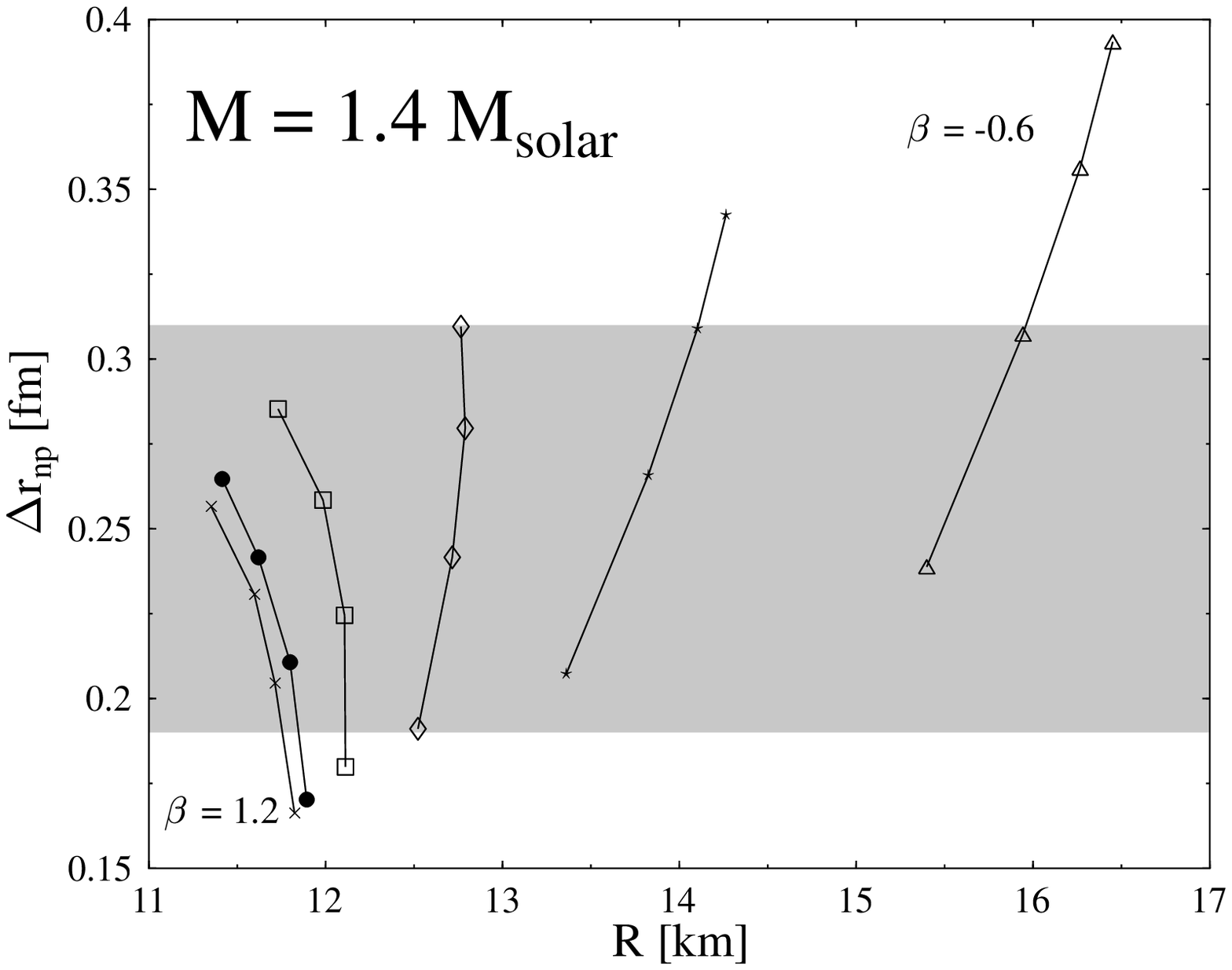}
\caption{\label{rpnbr}Relation of the difference  of neutron and proton radius $\Delta r_{np}$
of $^{208}$Pb (full symbols)
and the neutron star radius ($M = 1.4~M_{solar}$).
The coupling strengths $\beta$ and $r_\sigma$ are varied. Results for different
values of $\beta$ are marked with different symbols with values of 1.2, 1.0 0.6, 0.2, -0.2, -0.6
from the
left to the right. For every value of $\beta$ we varied $r_\sigma$ between 0 and 0.3,  with
large values of $r_\sigma$ yielding small values or $\Delta r_{np}$ as can also be seen
in Fig. \ref{rpnall}.
}
\end{figure}

A complete set of correlations between $\Delta r_{np}$ and $R_{ns}$ is given in
Fig.~\ref{rpnbr}. The various lines are results for different choices of $\beta$ and
show the dependence on the scalar-vector meson coupling $r_\sigma$.
The result shows that the information contained in $\beta$ and $r_\sigma$ is
rather complementary. Whereas the star radius largely depends on $\beta$ and little
on $r_\sigma$ in the case of the neutron skin the result is the opposite, a strong
$r_\sigma$ dependence and a small influence from the $\beta$ value.
However, as can be seen from the hatched area a 1\% measurement of the neutron
radius of lead covers about the whole range of results except for very small $\beta$
and large $r_\sigma$ values which are largely excluded by the results for
the asymmetry energy.

\subsection{Additional tests}

\begin{figure}
\includegraphics[width = 12cm]{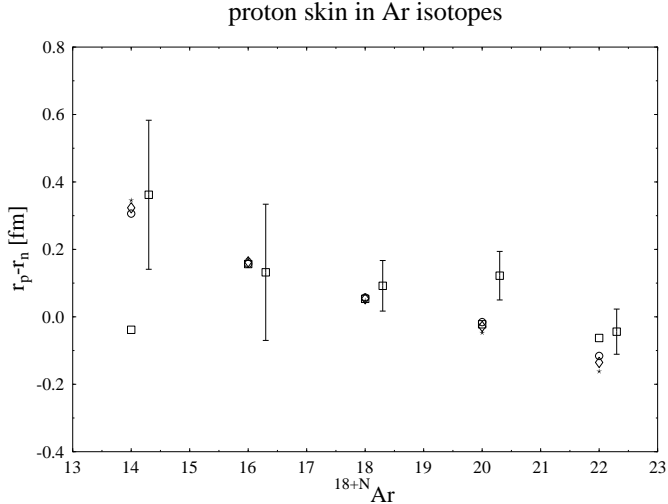}
\caption{\label{pskin}Proton skin for various Argon isotopes in comparison to experimental values
\cite{exp2}.
Results for different nonlinear coupling strengths are shown, the squares, circles, diamonds and
stars mark results for $\beta = 1.4, 1.0, -0.2, -0.6$, respectively.
}
\end{figure}

As a cross check to our discussion
we looked at the proton skin in different isotopes of argon. The skin was
experimentally deduced by combining measurements of heavy-ion reaction cross sections\cite{exp2}
with atomic isotope shift data \cite{exp1}.
The resulting values, with rather large error bars, are shown in Figure \ref{pskin}, compared
to the theoretical calculations using the same model for several values of the nonlinear
coupling varying $\beta$ ($r_\sigma$ is set to 0).
One can see that with the exception of the point at the high-end of coupling strengths for
the $^{32}Ar$ isotope the agreement with data is quite good, independent of the $\beta$ parameter.
This gives some confidence in the analysis that a change of the coupling does not automatically
lead to distorted results of other isospin-related observables.

\begin{figure}
\includegraphics[width = 10cm]{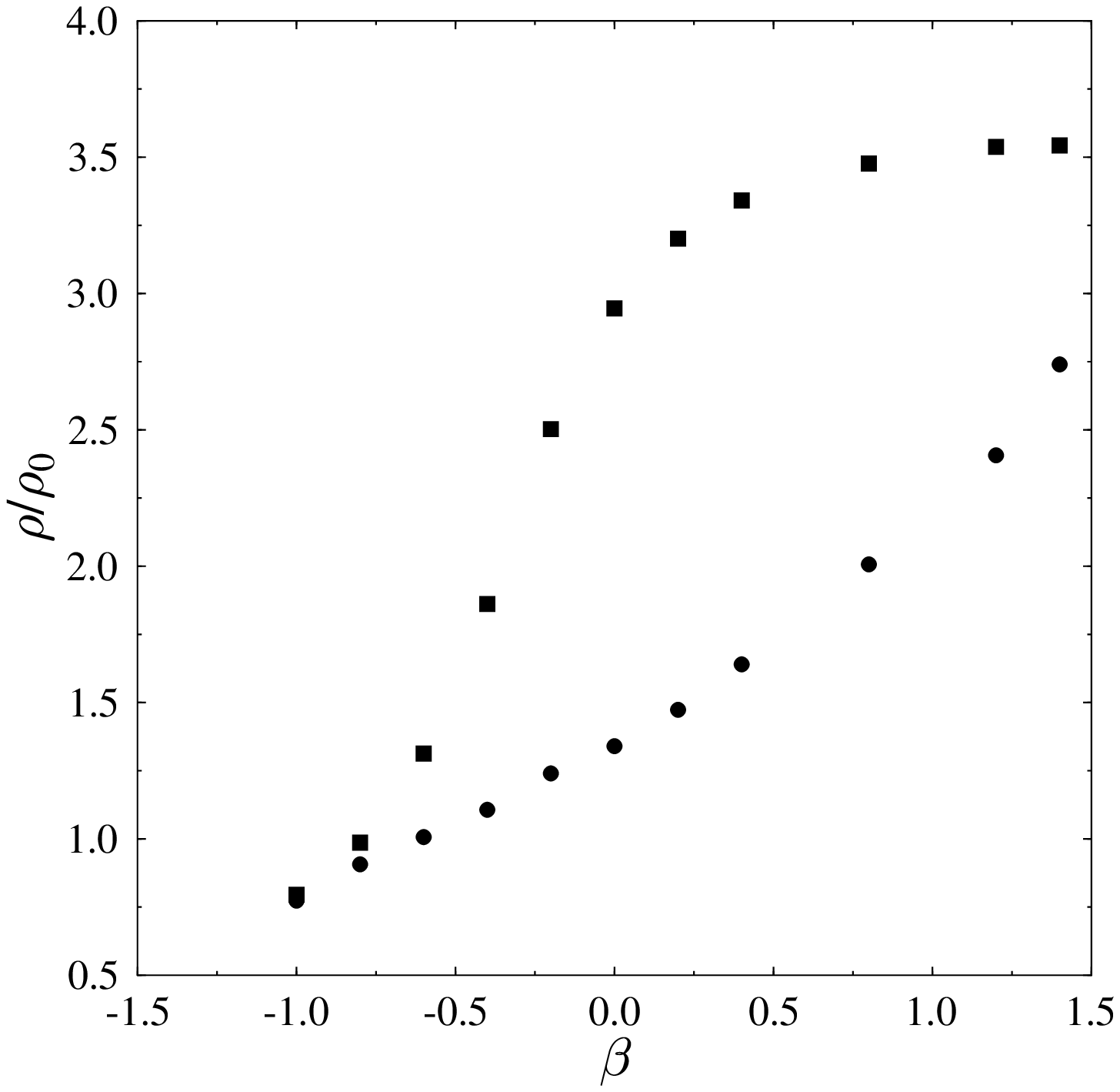}
\caption{\label{urca}Critical density for the onset of the URCA process for different nonlinear
coupling strengths $\beta$ setting $r_\sigma=0$. The squares show the maximum density for a neutron star with a typical mass
of $M = 1.4~M_{solar}$. The circles represent the minimum density necessary for the URCA process.
}
\end{figure}

Finally, in \cite{hor3} the importance of the URCA process for neutron star cooling was discussed
in this context. The URCA process is thought to be an important mechanism
for cooling the neutron star
after the supernova explosion at reasonably fast rates in agreement with astronomical observations.
This is done via the process $n \rightarrow p + e^- + \bar{\nu}_e$ where the neutrino escapes and
carries energy away from the system. This process can proceed when the fermi momenta of the
particles involved match :
\be
\label{ur}
p^F_n \leq p^F_p + p^F_e
\ee
Thus for the URCA process to take place one needs a significant amount of protons of roughly
$11 \%$ in the system. As the proton to neutron ratio in the star is influenced by the
isospin-dependence of the nuclear forces one has to check whether changing the force
might switch off the URCA process. We again consider a 1.4 solar mass star and compare the
central density of the star with the minimum density for the URCA process to occur according
to equation \ref{ur}. We see from Figure \ref{urca} that the minimum density is always reached.
Thus the discussion of the correlation of the star radius and neutron skin is not affected
by additional constraints from neutron star cooling considerations.

\section{Conclusion}
We studied the influence of couplings between vector and scalar mesons as well as
nonlinear vector-meson interactions and their influence on isospin-related observables
following the discussion in \cite{hor2} (see also \cite{bodmer}).
As basic model for our calculation we used a chiral
$\sigma-\omega$ type SU(3) model. This model has the
benefits of including those nonlinearities from the start and implements them
in a rather general SU(3) coupling scheme.
Our model has been successfully employed
in nuclear structure calculations and in calculations involving high-density and
high-temperature environments.
Since we also study neutron stars the benefit of having an SU(3) model is that hyperons
are automatically included in the model which is quite important
for getting reasonably small
neutron star masses in accordance with observation.
The results show clear dependence of the neutron skin in lead and barium on
couplings of vector and scalar mesons, whereas the neutron star radii depend largely
on non-linear terms in the vector-meson channel.
There is a distinct correlation of neutron star radii and neutron distributions in nuclei as
has already been observed in other nuclear structure models \cite{hor1,hor2}.
However, although the upcoming measurements of the lead radius using
parity-violating electron scattering experiments certainly will give additional
information for modelling the isospin-sector of the nuclear forces, it is likely
not stringent enough to nail down the parameters. Also possible measurements
of neutron star radii can only be useful in this regard if the accuracy
of the radius measurement is well below 1 km. At this level of radius uncertainty
also the detailed treatment of the hyperons in the neutron star enter, which
will make it very difficult to disentangle the different effects on the star radius.

To complete our investigation we have shown that the requirement of fast
neutron star cooling
does not constrain this discussion. Also
the measured value of the proton skins of argon isotopes
can be reproduced and are not strongly affected by varying the nonlinear couplings.

\begin{acknowledgments}
This work was supported by the U.S. Department of Energy, Nuclear
Physics Division (Contract No. W-31-109-Eng-38).
\end{acknowledgments}

\end{document}